\documentclass[aps,preprint]{revtex4}
\usepackage{epsfig}
\usepackage{graphics}
\begin{document}
\title{Coupled channel approach to the structure of the $X(3872)$}
\author{P.G. Ortega, J. Segovia, D.R. Entem, F. Fern\'{a}ndez.}
\affiliation{Grupo de F\'{\i}sica Nuclear and IUFFyM,
Universidad de Salamanca, E-37008 Salamanca,
Spain}
\vspace*{1cm} 
\begin{abstract}
We have performed a coupled channel calculation of the $1^{++}$ $c\bar c$ 
sector 
including $q\bar q$ and $DD^*$ molecular configurations. The calculation 
was done within a constituent quark model which successfully describes the meson 
spectrum, in particular the $c\bar c$ $1^{--}$ sector. 
Two and four quark configurations
are coupled using the $^3P_0$ model. 

The elusive $X(3872)$ meson appears as a new state with a high probability
for the $DD^*$ molecular component. When the mass difference between neutral and
charged states is included a large $D^0 {D^*}^0$ component is found which dominates
for large distances and breaks isospin symmetry in the physical state. 
The original $c\bar c(2 ^3P_1)$ state acquires a sizable $DD^*$ component and
can be identified with the $X(3940)$. 
We study the $B\to K \pi^+ \pi^-J/\psi$ and
$B\to K D^0 D^{*0}$ decays finding a good agreement with Belle 
and BaBar experimental data.
\vspace*{2cm}
\newline
\noindent Pacs: 12.39.-x, 13.25.Gv, 14.40.Gx.
\end{abstract}
\maketitle
\newpage
\section{Introduction.}
In the last years a number of exciting discoveries of new hadron states have 
challenged our description of the hadron spectroscopy. One of the most 
mysterious states is the well established $X(3872)$. It was first discovered by 
the Belle Collaboration in the $J/\psi \pi \pi$ invariant mass spectrum of the
decay $B^+\rightarrow K^+ \pi^+\pi^-J/\psi$~\cite{r1}. Its existence was soon 
confirmed by BaBar~\cite{r2}, CDF~\cite{r3} and D0~\cite{r4} Collaborations.
The world average mass is $M_X=3871.2\pm 0.5\,MeV$ and its width 
$\Gamma_X<2.3\,MeV$.
The measurements of the $X(3872)\rightarrow \gamma J/\psi$ decay~\cite{r5,r6} 
implies an even $C$-parity. Moreover angular correlation between final 
state particles in the $X(3872)\rightarrow \pi^+\pi^-J/\psi$ decay measured by 
Belle~\cite{r5} suggests that the $J^{PC}=0^{++}$ and $J^{PC}=0^{+-}$ may be 
ruled out and strongly favors the $J^{PC}=1^{++}$ quantum numbers although the 
$2^{++}$ combination cannot be excluded. A later analysis by CDF 
Collaboration~\cite{r8} of the same decay is compatible with the Belle results 
and concludes from the dipion mass spectrum that the most likely quantum numbers 
should be $J^{PC}=1^{++}$ but cannot totally exclude the $J^{PC}=2^{-+}$ 
combination . 
These conclusions were confirmed by a new CDF analysis of the decay 
$X(3872)\to \pi^+ \pi^- J/\psi$ followed by $J/\psi \to \mu^+ \mu^-$
excluding all the other possible quantum numbers at $99.7\,\%$ 
confidence level~\cite{m8}.
However the small phase space available for the decay 
$X(3872)\rightarrow D^0\bar D^0\pi ^0$ observed by Belle~\cite{r9} discards the 
$J=2$ leaving the $1^{++}$ assignment as the most probable option. 

In the $1^{++}$ sector the only well established state in the PDG~\cite{pdg} 
is the $\chi_{c_1}(1P)$ with a mass $M=3510.66\pm0.07\,MeV$. The first
excitation is expected around $3950\,MeV$. In this energy region
Belle has reported the observation of three resonant structures denoted
by $X(3940)$, $Y(3940)$ and $Z(3930)$. The last one was observed by Belle
in the $\gamma \gamma \to D\bar D$ reaction~\cite{m9a} and is already
included in the PDG as the $\chi_{c_2}(2P)$. The $X(3940)$ has been
seen as a peak in the recoiling mass spectrum of $J/\psi$ produced
in $e^+e^-$ collision. Its main decay channel is $DD^*$~\cite{m9b}.
The $Y(3940)$ appears as a threshold enhancement in the $J/\psi \omega$
invariant mass distribution of the $B\to J/\psi \omega K$ decay~\cite{m9c}.

The relative decay rates outlines a puzzling structure for the $X(3872)$.
The $\gamma J/\psi$ and $\gamma \psi'$ decay rates~\cite{m8b}
\begin{eqnarray}
\frac{X(3872)\rightarrow \gamma J/\psi}{X(3872)\rightarrow 
\pi^+\pi^-J/\psi}&=&0.33\pm0.12
\nonumber \\
\frac{X(3872)\rightarrow \gamma\psi'}{X(3872)\rightarrow \pi^+\pi^-J/\psi}&=&1.1\pm0.4
\label{mes-w}
\end{eqnarray}
suggest a $c\bar c$ structure whereas the $X(3872)\to \pi^+ \pi^- \pi^0 J/\psi$
decay mode 
\begin{eqnarray}
\frac{X(3872)\rightarrow \pi^+\pi^-\pi^0J/\psi}{X(3872)\rightarrow 
\pi^+\pi^-J/\psi}&=&1.0\pm0.4\pm 0.3
\end{eqnarray}
indicates a very different one~\cite{m9}. The dipion mass spectrum in the
$\pi^+ \pi^- J/\psi$ channel shows that the pions come from the $\rho^0$
resonance. On the other hand the $\pi^+\pi^-\pi^0$ mass spectrum 
has a strong peak around
$750\,MeV$ suggesting that the process is dominated by a $\omega$
meson. Thus the ratio $R\sim1$ indicates that there should be an isospin violation incompatible with a traditional charmonium assumption.

Concerning the mass value, in 2006 Belle measured~\cite{r9} an enhancement
in the $D^0 D^0 \pi^0$ channel just above the $D^0 D^{*0}$ threshold
using the $B^+ \to K^+ D^0 D^0 \pi^0$ decay. The amazing aspect of
this enhancement is that it appears at $M_X=3875.2\pm0.7^{+0.3}_{-1.6}\pm0.8\,MeV$
just $3\,MeV$ above the $M_X$ world average mass value. This fact
triggered a new discussion about the possibility of two different
charmonium like states. The Belle mass value was
confirmed later by the BaBar Collaboration~\cite{BaBarDDD}.
Last year the Belle Collaboration announced a new 
measurement of the $B \to K D^0 D^0 \pi^0$ decay~\cite{BelleDDD}
with a lower position of the $X(3872)$ peak in $M_X=3872.6^{+0.5}_{-0.4}\pm 0.4\,MeV$.
New data of the $\pi^+\pi^-J/\Psi$ decay has been also recently reported by the
Belle~\cite{BelleDJPsi}, BaBar~\cite{BaBarDJPsi} and CDF~\cite{CDFDJPsi} Collaborations,
confirming a mass value in agreement with the world average.

The $X(3872)$ mass is difficult to reproduce by the standard quark models (see 
Ref.~\cite{r10} for a review). The state appears to be too heavy for a $1D$ 
charmonium state and too light for a $2P$ charmonium one. Moreover no 
four-quark bound state configurations have been found in this mass region which 
rules out the possibility that this particle was a compact tetraquark 
system~\cite{r11,m11}.

An important property of the $X(3872)$ is that its mass is extremely close to 
the $D^0 {D^*}^0$ threshold with a difference using PDG values 
given by $-0.6\pm0.6\,MeV$. The proximity of the 
$D^0{D^*}^0$ threshold made the $X(3872)$ a natural candidate to a 
$C=+$ $D^0{D^*}^0$ molecule. 
The hypothesis of a $DD^*$ molecule mainly bound by pion exchange has been
suggested by several authors~\cite{r12}. In particular, in Ref.~\cite{r13}
it is argued that the $X(3872)$ is a $J^{PC}=1^{++}$ $D^0{D^*}^0$
molecule stabilized by admixture of $\rho J/\psi$ and $\omega J/\psi$
states. The author shows that pion exchange alone can not bind the molecule
being the combined effect of pion exchange and coupled channels responsible
for that. The $D^0 {D^*}^0$ component dominates the wave function at
the experimental binding becoming all other contributions small.

The molecular interpretation runs into trouble when it tries to explain
the high $\gamma \psi'$ decay rate. For a molecular state this can be only proceed
through annihilation diagrams and hence is very small.

This puzzling situation suggests for the $X(3872)$ state a combination of a $2P$ 
$c\bar c$ state and a weakly-bound 
$D^0{D^*}^0$ molecule~\cite{m9c,m8b}. 
The experimental assignment 
$J^{PC}=1^{++}$ favors this conclusion because it allows the molecule to be in a 
relative $S$-wave state whereas the corresponding $c\bar c$ should be in a 
relative $P$-wave state. Then the masses of the additional light quarks are 
compensated by the angular momentum excitation and both configurations may be 
almost in the same mass region. Similar behavior has been already observed in the open
charmed sector~\cite{m10}. 
Recently Zhang {\it et al.}~\cite{r15} have analyzed, using the coupled 
channel Flatt\'e formula, the 
$B\to K D^0 D^{0}\pi^0$~\cite{BelleDDD}  and
$B\to K \pi^+ \pi^-J/\Psi$~\cite{BelleDJPsi} 
Belle data.
They found that a third sheet pole close,
but below, the $D^0{D^*}^0$ threshold is needed to describe the data, which 
supports the idea of the $X(3872)$ as a mixed state of $\chi'_{c1}$ and 
$D^0{D^*}^0$ components. An updated Flatt\'e analysis of the same data
together with the new BaBar data of the same reactions~\cite{BaBarDJPsi,BaBarDDD}
has been performed in Ref.~\cite{Kala1} assuming a mechanism for the
$X(3872)$ production via the charmonium components. The authors
conclude that the data clearly indicates a sizable $c\bar c$ $2\,^3P_1$
component in the $X(3872)$ wave function. Finally Dong {\it et al.}~\cite{g1} show in their analysis of the
$J/\psi ~\gamma$ and $\psi(2S)~ \gamma$ decay modes of the $X(3872)$ that the large value of the ratio
$BR(X(3872)\rightarrow J/\psi \gamma)/BR(X(3872)\rightarrow \psi(2S) \gamma)$ measured by the BaBar Collaboration provide a constraint on the value of the $c\bar c$ component in the $X(3872)$. From the experimental values they  deduce a small admixture of the $c\bar c$ component

Having in mind these evidences, in this paper we perform
a microscopic coupled channel calculation of the 
$1^{++}$ sector including both $c\bar c$ and $DD^*$ 
states. The calculation is done in the framework of a constituent quark model
widely used in hadronic spectroscopy. The paper is organized as follows. In the next section we review 
the main ingredients of our model. Section III is devoted to discuss
the numerical procedures and the results . Finally we summarize the main findings of our work in the last 
section.
\section{The Model.}
\subsection{The constituent quark model}

The constituent quark model used in this work has been extensively
described elsewhere ~\cite{r16,r16b} and therefore 
we will only summarize here its most
relevant aspects. The model is based on the assumption that the light 
constituent quark mass appears as a consequence of the spontaneous breaking of 
the chiral symmetry at some momentum scale. As a consequence the quark 
propagator gets modified and quarks acquire a dynamical momentum dependent 
mass. The simplest Lagrangian 
must therefore contain chiral fields to compensate the mass term and can be expressed as  ~\cite{r17}
 
\begin{equation}\label{lagrangian}
{\mathcal L}
=\overline{\psi }(i\, {\slash\!\!\! \partial} -M(q^{2})U^{\gamma_{5}})\,\psi 
\end{equation}
where $U^{\gamma _{5}}=\exp (i\pi ^{a}\lambda ^{a}\gamma _{5}/f_{\pi })$,
$\pi ^{a}$ denotes nine pseudoscalar fields $(\eta _{0,}\vec{\pi }
,K_{i},\eta _{8})$ with $i=$1,...,4 and $M(q^2)$ is the constituent mass.
This constituent quark mass, which vanishes at large momenta and is frozen at low 
momenta at a value around 300 MeV, can be explicitly obtained from the theory but
its theoretical behavior can be simulated by parameterizing 
$M(q^{2})=m_{q}F(q^{2})$ where $m_{q}\simeq $ 300 MeV, and

\begin{equation}
F(q^{2})=\left[ \frac{{\Lambda}^{2}}{\Lambda ^{2}+q^{2}}
\right] ^{\frac{1}{2}} \, .
\end{equation} 
The cut-off $\Lambda$ fixes the
chiral symmetry breaking scale.

The Goldstone boson field matrix $U^{\gamma _{5}}$ can be expanded in terms of boson fields,

\begin{equation}
U^{\gamma _{5}}=1+\frac{i}{f_{\pi }}\gamma ^{5}\lambda ^{a}\pi ^{a}-\frac{1}{%
2f_{\pi }^{2}}\pi ^{a}\pi ^{a}+...
\end{equation}
The first term of the expansion generates the constituent quark mass while the
second gives rise to a one-boson exchange interaction between quarks. The
main contribution of the third term comes from the two-pion exchange which
has been simulated by means of a scalar exchange potential.

In the heavy quark sector 
chiral symmetry is explicitly broken and this type of interaction does not act. 
However it constrains the model parameters through the light meson phenomenology 
and provides a natural way to incorporate the pion exchange interaction in the 
$DD^*$ dynamics. 

Beyond the chiral symmetry breaking scale one 
expects the dynamics to be governed by QCD perturbative effects.
They are taken 
into account through the one gluon-exchange interaction \cite{OGE} derived from 
the lagrangian 

\begin{equation}
\label{Lg}
{\mathcal L}_{gqq}=
i{\sqrt{4\pi\alpha _{s}}}\, \overline{\psi }\gamma _{\mu }G^{\mu
}_c \lambda _{c}\psi  \, ,
\end{equation}
where $\lambda _{c}$ are the SU(3) color generators and G$^{\mu }_c$ the
gluon field. 

The other QCD nonperturbative effect corresponds to confinement,
which prevents from having colored hadrons.
Such a term can be physically interpreted in a picture in which
the quark and the antiquark are linked by a one-dimensional color flux-tube.
The spontaneous creation of light-quark pairs may
give rise at same scale to a breakup of the color flux-tube \cite{Bali}. This can be translated
into a screened potential~\cite{c28} in such a way that the potential
saturates at the same interquark distance.
\begin{equation}
V_{CON}(\vec{r}_{ij})=\{-a_{c}\,(1-e^{-\mu_c\,r_{ij}})+ \Delta\}(\vec{%
\lambda^c}_{i}\cdot \vec{ \lambda^c}_{j})\,
\end{equation}

\noindent where $\Delta $ is a global constant to fit the origin of
energies. At short distances this potential presents a linear behavior with
an effective confinement strength $a=-a_{c}\,\mu _{c}\,(\vec{\lambda ^{c}}%
_{i}\cdot \vec{\lambda ^{c}}_{j})$ while it becomes constant at large
distances. It has been shown that this form of the 
potential is important to explain the huge degeneracy observed in the high 
excited light meson spectrum~\cite{r18} and turns out to be very important for 
the correct assignment of $J^{PC}=1^{--}$ charmonium states~\cite{r19}. Explicit 
expressions for all these interactions are given in~\cite{r19}.

All these ingredients are needed to explain the hadronic phenomenology. Apart from the obvious confinement potential, gluon exchange is demanded from the hyperfine splitting in charmonium. Moreover, pion exchange is one of the best established interaction in nature being its parameters constraint by a huge amount of experiments. When Goldstone boson exchanges are considered at the quark level together with the OGE the possibility of double counting emerges. This problem has been studied in the literature concluding that the pion can be safely exchanged together with the gluon ~\cite{n16}.

Constituent quark models are also criticized because they only
incorporate a limited sector of the Fock space. In particular its
applicability to high excited states may be questionable as more
thresholds open up. In our case the parameters of the model has been
fixed in the low lying part of the spectrum where these effects are
more easily incorporated into them. Furthermore the main
contribution of the open channels are taken into account
by the screened confinement potential.

\subsection{The coupled channel approach}
To model the $1^{++}$ $c\bar c$ system
we assume that the hadronic state is
\begin{equation} \label{ec:funonda}
 | \Psi \rangle = \sum_\alpha c_\alpha | \psi_\alpha \rangle
 + \sum_\beta \chi_\beta(P) |\phi_{M_1} \phi_{M_2} \beta \rangle
\end{equation}
where $|\psi_\alpha\rangle$ are $c\bar c$ eigenstates of the two body
Hamiltonian, 
$\phi_{M_i}$ are $c\bar n$ ($\bar c n$) eigenstates describing 
the $D$ ($\bar D$) mesons, 
$|\phi_{M_1} \phi_{M_2} \beta \rangle$ is the two meson state with $\beta$ quantum
numbers coupled to total $J^{PC}$ quantum numbers
and $\chi_\beta(P)$ is the relative wave 
function between the two mesons in the molecule. As we always work with eigenstates
of the $C$-parity operator we use the usual notation in which $DD^*$ is the right
combination of $D\bar D^*$ and $D^* \bar D$.

The coupling between the two sectors requires the creation of a light quark pair $n\bar n$. Similar to
the strong decay process this coupling should be in principle driven by the same interquark hamiltonian which determines the spectrum. However Ackleh et al.~\cite{n17} have shown that the quark pair creation $^3P_0$ 
model~\cite{r21}, gives  similar results to the microscopic calculation. The model assumes that the pair creation Hamiltonian is 
\begin{equation} 
\mathcal{H}=g \int d^3x \,\, \bar \psi(x) \psi(x)
\end{equation}
which in the non-relativistic reduction is equivalent to the transition
operator~\cite{Bonnaz}
\begin{eqnarray}
T&=&-3\sqrt{2}\gamma'\sum_\mu \int d^3 p d^3p' \,\delta^{(3)}(p+p')
\left[ \mathcal Y_1\left(\frac{p-p'}{2}\right) b_\mu^\dagger(p)
d_\nu^\dagger(p') \right]^{C=1,I=0,S=1,J=0}
\label{TBon}
\end{eqnarray}
where $\mu$ ($\nu=\bar \mu$) are the quark (antiquark) quantum numbers and
$\gamma'=2^{5/2} \pi^{1/2}\gamma$ with $\gamma= \frac{g}{2m}$ is a dimensionless constant 
that gives the strength of 
the $q\bar q$ pair creation from the vacuum.
From this operator we define the transition
potential $V_{\beta \alpha}(P)$ within the $^3 P_0$ model as~\cite{Kala0} 

\begin{equation}
\label{Vab}
	\langle \phi_{M_1} \phi_{M_2} \beta | T | \psi_\alpha \rangle =
	P \, V_{\beta \alpha}(P) \,\delta^{(3)}(\vec P_{\mbox{cm}})
\end{equation}
where $P$ is the relative momentum of the two meson state.

Using the wave-function from Eq. (\ref{ec:funonda}) and the coupling
Eq. (\ref{Vab}) we arrive to the coupled equations
\begin{eqnarray}
M_\alpha \,c_\alpha + \sum_\beta\int V_{\alpha \beta}(P) \chi_{\beta} (P)\,P^2 \,dP &=& E \,c_\alpha 
\nonumber \\
\sum_{\beta} \int H^{M_1 M_2}_{\beta'\beta} (P',P) 
\chi_\beta(P) \, P^2 \, dP + \sum_\alpha V_{\beta' \alpha}(P') c_\alpha &=& E \,\chi_{\beta'}(P')
\label{coupleE}
\end{eqnarray}
where $M_\alpha$ are the masses of the bare $c\bar c$ mesons and
$H^{M_1 M_2}_{\beta'\beta}$ is the RGM Hamiltonian for the two meson
states obtained from the $q\bar q$ interaction.

Solving the coupling with $c\bar c$ states
we finally end up with an Schr\"odinger type equation
for the relative wave function of the two meson state
\begin{equation}
	\sum_{\beta} \int \left( H^{M_1 M_2}_{\beta'\beta} (P',P) + 
	V^{eff}_{\beta'\beta} (P',P) \right)
	\chi_\beta(P) \, P^2 \, dP = E \,\chi_{\beta'}(P')
\end{equation}
where 
\begin{equation}
	V^{eff}_{\beta'\beta}(P',P) = 
	\sum_\alpha \frac{V_{\beta'\alpha}(P') V_{\alpha\beta}(P)}{E-M_\alpha}
\end{equation}
is an effective interaction between the two mesons due to the coupling
with intermediate $c\bar c$ states.

In this way we study the influence of the $c \bar c$ states on the dynamics
of the two meson states. This is a different point of view from the usually
found in the literature where the influence of two meson states (in general
without meson-meson interaction) in the mass and width of $c \bar c$ states
is studied~\cite{Kala0}. Our approach allows to generate new states through the meson-meson
interaction due to the coupling with $c\bar c$ states and to the underlying
$q\bar q$ interaction. As we will see the renormalization effects of the 
$c\bar c$ mass due to this channel is small.

The $c\bar c$ probabilities are given by
\begin{equation}
	c_\alpha = \frac{1}{E-M_\alpha}
	\sum_\beta\int V_{\alpha\beta} (P) \chi_\beta(P) P^2 \,dP
\end{equation}
with the normalization condition 
$1=\sum_\alpha |c_\alpha|^2 + \sum_\beta \langle \chi_\beta | \chi_\beta \rangle$.

\subsection{Flatt\'e parametrization}
In order to compare the predictions of our model with the recent Belle and BaBar experimental
data we obtain from Eq. (\ref{coupleE}) a Flatt\'e-like parametrization
of the $DD^*$ near threshold amplitude following Ref.~\cite{Baru}. We remind here
the main ideas.

From Eq. (\ref{coupleE}), and neglecting the $DD^*$ interaction, 
one can easily derive the $DD^*$ scattering
amplitude
\begin{eqnarray}
	F_{DD^*}^\beta (P,P;E) &=& -\pi\mu\sum_\alpha \frac{V_{\beta\alpha}^2(P)}
	{E-M_\alpha+g_{DD^*}^\alpha(E)}
\end{eqnarray}
where 
the function $g_{DD^*}^\alpha(E)$ is given by
\begin{eqnarray}
	g_{DD^*}^\alpha(E) &=& \sum_\beta \int \frac{V_{\beta\alpha}^2(P)}{\frac{P^2}{2\mu}+M_D+M_{D^*}-E-i0^+} \,P^2\,dP.
\end{eqnarray}
For small binding energies $\epsilon = M_D + M_{D^*}-E$ it can be expanded as
\begin{eqnarray}
	g_{DD^*}^\alpha(E) &=& \bar E_{DD^*}^\alpha + \frac i 2 \Gamma_{DD^*}^\alpha + \mathcal{O}(4\mu^2\epsilon/\Lambda^2)
\end{eqnarray}
where
\begin{eqnarray}
	\bar E_{DD^*}^\alpha &=& 2 \mu \sum_\beta \int_0^\infty V_{\beta\alpha}^2(P) \,dP
	\\
	\Gamma_{DD^*}^\alpha &=& 2\pi \mu \sum_\beta V_{\beta\alpha}^2(0) \,P
\end{eqnarray}
and $\Lambda\gg \epsilon$ is the characteristic scale of the $V_{\alpha \beta}$ production
amplitude 
which may correspond to
the scale of the quark wave function and it's assume to be much bigger than the
binding energy of the physical state.

A straightforward generalization to include the $DD^*$ charged states and
other channels gives the expression for the near threshold $DD^*$ scattering amplitude 
\begin{eqnarray}
	F_{DD^*} = -\frac{1}{2P} \frac{\Gamma_{DD^*}}{E-E_f+\frac i 2(\Gamma_{D^0D^{*0}}+
	\Gamma_{D^+D^{*-}}+\Gamma(E))+\mathcal{O}(4\mu^2\epsilon/\Lambda^2)}
\label{flate}
\end{eqnarray}
where 
$\Gamma(E)$ accounts for the width due to other processes different from
the opening of the near $DD^*$ threshold. Eq. (\ref{flate})
corresponds to a Flatt\'e parametrization with
\begin{eqnarray}
	D(E) &=& E-E_f 
	+\frac i 2(\Gamma_{D^0D^{*0}}+
	\Gamma_{D^+D^{*-}}+\Gamma(E))+\mathcal{O}(4\mu^2\epsilon/\Lambda^2).
\end{eqnarray}
Now assuming, as in Ref.~\cite{Kala1}, that the short range dynamics of the weak
$B\to KX(3872)$ transition can be absorbed into a coefficient $\mathcal{B}$
we are able to write the differential rates in the Flatt\'e 
approximation as
\begin{eqnarray}
	\frac{d Br(B\to K D^0 D^{*0})}{dE} &=&
	\mathcal{B} 
	\frac{1}{2\pi} 
	\frac{\Gamma_{D^0D^{*0}}(E)}{|D(E)|^2}.
\end{eqnarray}

The analysis of the $B\to KX(3872) \to K \pi^+\pi^- J/\psi$ data is more 
involved because we have to calculate the $DD^*\to \pi^+\pi^-J/\psi$
transition amplitude.

\begin{figure}[t]
\begin{center}
\scalebox{1.0}{\includegraphics{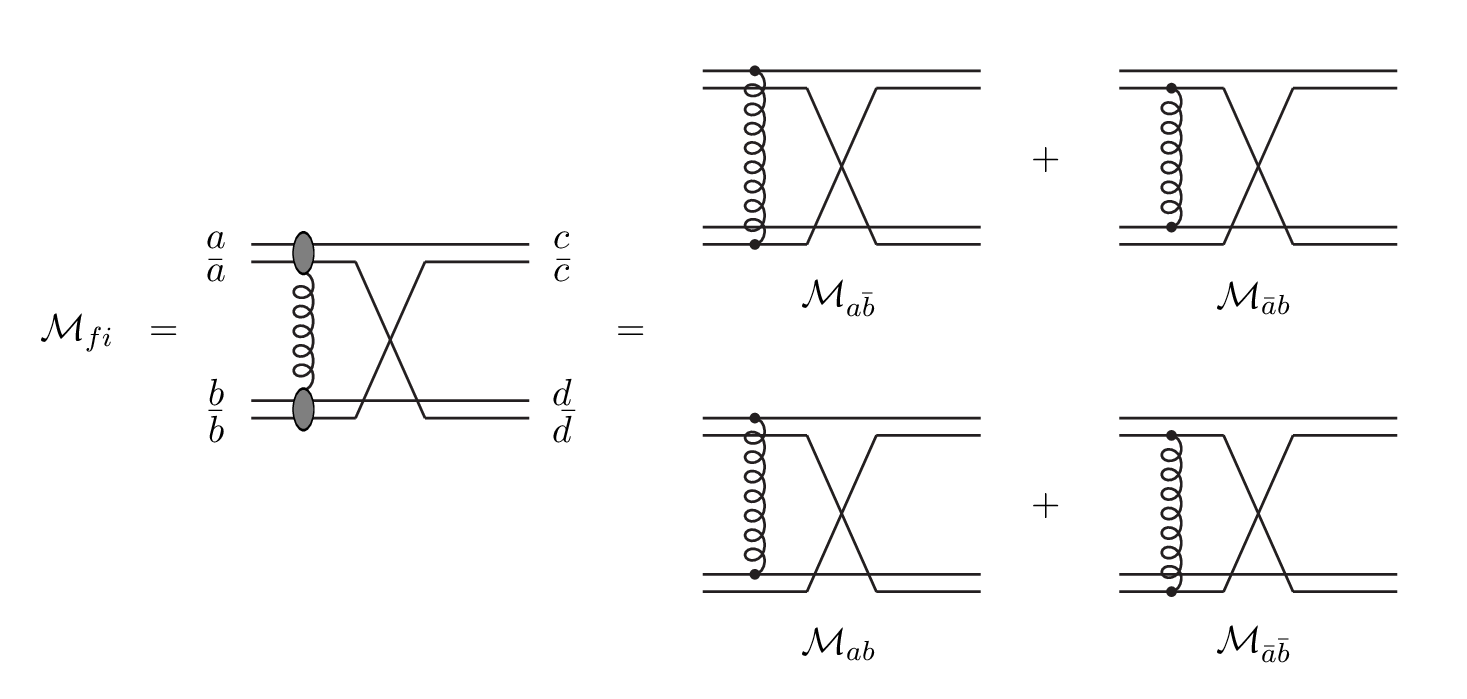}}
\caption{\label{diag} Diagrams included in the quark rearrangement
process $DD^* \to \rho J/\psi$.}
\end{center}
\end{figure}

This can consistently be done in our formalism assuming that the process
takes place through the $DD^*$ components of the $X(3872)$ which decays
in $\rho J/\psi$ and then into the final $\pi^+\pi^-J/\psi$ states.
The decay width of the process is given by
\begin{eqnarray}
	\Gamma_{\pi^+\pi^-J/\psi} =
	\sum_{JL} \int_0^{k_{max}} dk 
	\frac{\Gamma_{\rho}}{(M_X-E_\rho-E_{J/\psi})^2+\frac{\Gamma_\rho^2}{4}}
	\left|\mathcal{M}^{JL}_{X\to \rho J/\psi}(k)\right|^2.
\end{eqnarray}
The amplitude $\mathcal{M}^{JL}_{X\to \rho J/\psi}$ is calculated
in our model by the rearrangement diagrams of Fig.~\ref{diag},
averaged with the $DD^*$ component of the $X(3872)$ wave function.
The rearrangement diagrams are calculated following Ref.~\cite{BarnesSwanson}.
The amplitude is given by
\begin{eqnarray}
	\mathcal{M}_{fi} &=& \sum_{i=a,\bar a;j=b,\bar b} \mathcal{M}_{ij}
\end{eqnarray}
where
\begin{eqnarray}
	\mathcal{M}_{ij}(\vec P',\vec P) &=& 
	\langle \phi_{M'_1} \phi_{M'_2}|H_{ij}^O| \phi_{M_1} \phi_{M_2} \rangle
	\langle \xi_{M'_1M'_2}^{SFC} |\mathcal{O}_{ij}^{SFC}| \xi_{M_1M_2}^{SFC} \rangle
\end{eqnarray}
and the orbital part can be written as (e.g. for the case $(ij)=(a\bar b)$)
\begin{eqnarray}
	\langle \phi_{M'_1} \phi_{M'_2}|H_{ij}^O| \phi_{M_1} \phi_{M_2} \rangle &=&
	\int d^3 P_{M'_1}d^3 P_{M'_2}d^3 P_{M_1}d^3 P_{M_2} \,\,
	\phi^*_{M'_1}(P_{M'_1}) \phi^*_{M'_2}(P_{M'_2}) 
	\delta(\vec{P}_{M'_2}-\vec{P}_{M_1})
	\nonumber \\ &&
	\delta(\vec{P}_{M'_2}-\vec{P}_{M_2}-(\vec{P}'-\vec{P}))
	H(-\frac 1 2(\vec{P}_{M_1}+\vec{P}_{M_2})+\vec{P}_{M'_1}
	+\frac 1 2(\vec{P}'-\vec{P}))
	\nonumber \\ &&
	\phi_{M_1}(P_{M_1}) \phi_{M_2}(P_{M_2}).
\end{eqnarray}
The spin-flavor-color matrix elements are taken from
Ref.~\cite{BarnesSwanson}. 

Once the decay width $\Gamma_{\pi^+\pi^-J/\psi}$ is calculated, the 
differential rate is given by
\begin{eqnarray}
	\frac{d Br(B\to K  \pi^+\pi^-J/\psi)}{dE} &=&
	\mathcal{B} 
	\frac{1}{2\pi} \frac{\Gamma_{\pi^+\pi^-J/\psi}(E)}{|D(E)|^2}.
\end{eqnarray}
In order
to compare with the experimental data
we determine the number of events distributions from
the differential cross section
\begin{eqnarray}
	N_{Belle}^{\pi\pi J/\psi} (E) &=&
	2.5 \mbox{[MeV]} \, \left( \frac{131}{8.3\, 10^{-6}}\right) 
	\frac{d Br(B\to K  \pi^+\pi^-J/\psi)}{dE}
	\\
	N_{Belle}^{D^0\bar D^0\pi^0} (E) &=&
	2.0 \mbox{[MeV]} \, \left( \frac{48.3}{0.73\, 10^{-4}}\right) 
	\frac{d Br(B\to K  D^0\bar D^0\pi^0)}{dE}
	\\
	N_{BaBar}^{\pi\pi J/\psi} (E) &=&
	5 \mbox{[MeV]} \, \left( \frac{93.4}{8.4\, 10^{-6}}\right) 
	\frac{d Br(B\to K  \pi^+\pi^-J/\psi)}{dE}
	\\
	N_{BaBar}^{D^0 D^{*0}} (E) &=&
	2.0 \mbox{[MeV]} \, \left( \frac{33.1}{1.67\, 10^{-4}}\right) 
	\frac{d Br(B\to K  D^0\bar D^{*0})}{dE}.
\end{eqnarray}
In all reactions a background is taken into account modelled
as in Ref.~\cite{Kala1}.
For the $B\to K  D^0\bar D^0\pi^0$ the $D^{0}D^{*0}$ signal
interferes with the background and so a phase $\phi^{Belle}=0^0$
and $\phi^{BaBar}=324^0$
have been introduced. Also the
experimental branching ratio $B(D^{*0}\to D^0\pi^0)=0.62$
is introduced. We use a value for $\mathcal{B}=3.5\,10^{-4}$ which
is in the order of the one used in Ref.~\cite{Kala1}.

\section{results}
\subsection{Numerical methods}
To found the quark-antiquark bound states  we solve the 
Schr\"odinger equation using the Gaussian Expansion Method~\cite{r20}. 
In this method the radial wave functions solution of the Schr\"odinger equation 
are expanded in terms of basis functions
\begin{equation}
R_{\alpha}(r)=\sum_{n=1}^{n_{max}} b_{n}^\alpha \phi^G_{nl}(r)
\end{equation} 
where $\alpha$ refers to the channel quantum numbers.
The coefficients $b_{n}^\alpha$ and the eigenenergy $E$ are determined from the 
Rayleigh-Ritz variational principle
\begin{equation}
\sum_{n=1}^{n_{max}} \left[\left(T_{n'n}^\alpha-EN_{n'n}^\alpha\right)
b_{n}^\alpha+\sum_{\alpha'}
\ V_{n'n}^{\alpha\alpha'}b_{n}^{\alpha'}=0\right]
\end{equation}
where the operators $T_{n'n}^\alpha$ and $N_{n'n}^\alpha$ are diagonal and 
the only operator which mix the different channels is the potential 
$V_{n'n}^{\alpha\alpha'}$.
 
To solve the four body problem we also use the gaussian expansion of the two body wave functions obtained from the solution of the Schr\"odinger equation. This procedure allows us to introduce in  variational way possible distortions of the two body wave function within the molecule. Using these wave functions Eq. (13) reduces to a matrix equation by Gauss integration.

A crucial problem of the variational methods is how to choose the radial functions $\phi^G_{nl}(r)$ in order to have a minimal, but enough, number of basis functions. Following~\cite{r20} we employ gaussians trial functions whose ranges are in geometric progression. The geometric progression is useful in optimizing the ranges with a small number of free parameters. Moreover the distribution of the gaussian ranges in geometric progression is dense at small ranges, which is well suited for making the wave function correlate with short range potentials. The fast damping of the gaussian tail is not a real problem since we can choose the maximal range much longer than the hadronic size.
\subsection{Results}
The calculation is parameter free since
all the parameters are taken from the previous calculation~\cite{r16b, r19} including the $\gamma=0.26$ parameter in Eq. (\ref{TBon}).
This value was fitted to the reaction
$\psi(3770) \to D D$
which is the only well established charmonium
strong decay.
This way to determine the value of $\gamma$ might overestimate it
since the $\psi(3770)$ is very close to the $DD$ threshold and
FSI effects, which were not included, might be relevant~\cite{r22}.

We first perform an isospin symmetric calculation including $^3 S_1$ and
$^3D_1$ $DD^*$ partial waves and taking the $D$ and $D^*$ masses as
average of the experimental values between charged states.
If we neglect the coupling to $c\bar c$ states we don't get a
bound state for the $DD^*$ molecule in the $1^{++}$ channel,
neither in the $I=0$ nor in the $I=1$ channels. The interaction
coming from OPE is attractive in the $I=0$ channel but not enough to bind the system,
even allowing for distortion in the meson states.

Now we include in the $I=0$ channel the coupling to $c\bar c$ states. The most relevant are the $1^{++}$
ground
and first excited states with bare masses within the model given by
\begin{equation}
 \begin{array}{rcl}
  c\bar c(1^3P_1) \to \mbox{ }M=3503.9\mbox{ }MeV \\
  c\bar c(2^3P_1) \to \mbox{ }M=3947.4\mbox{ }MeV.
 \end{array}
\end{equation}

\begin{table}
\begin{center}
\begin{tabular}{cccccc}
\hline
\hline
 & $M\,(MeV)$ & $c\bar c(1 ^3P_1)$ & $c\bar c(2 ^3P_1)$ 
& $D^0{D^*}^0$ & $D^\pm{D^*}^\mp$ \\
\hline
   & 3936 & $ 0  \,\%$ & $79\,\%$   & $10.5\,\%$  & $10.5\,\%$ \\
 A & 3865 & $1\,\%$    & $32\,\%$   & $33.5\,\%$  & $33.5\,\%$ \\
   & 3467 & $95\,\%$   & $ 0  \,\%$ & $2.5\,\%$   & $2.5\,\%$  \\
\hline
   & 3937 & $   0\,\%$ & $ 79\,\%$   & $   7\,\%$  & $   14\,\%$ \\
 B & 3863 & $ 1  \,\%$ & $30\,\%$   & $46\,\%$    & $23\,\%$   \\
   & 3467 & $ 95\,\%$   & $   0\,\%$ & $  2.5\,\%$   & $  2.5\,\%$  \\
\hline
   & 3942 & $   0\,\%$ & $ 88\,\%$   & $   4\,\%$  & $   8\,\%$ \\
 C & 3871 & $ 0  \,\%$ & $ 7\,\%$   & $83\,\%$    & $10\,\%$   \\
   & 3484 & $ 97\,\%$   & $   0\,\%$ & $  1.5\,\%$   & $  1.5\,\%$  \\
\hline
\hline
\end{tabular}
\caption{\label{t1} Masses and channel probabilities for the three
states in three different calculations. The first three states are
found when we perform and isospin symmetric calculation with a value
of $\gamma$ fit to the decay $\psi(3770)\to DD$. The second three states
shows the effect of isospin breaking in the $DD^*$ masses. The last
three states correspond to a value of $\gamma=0.19$ that fits the 
experimental mass of the $X(3872)$. 
The probability is shown as zero when it is less than $0.5\,\%$.}
\end{center}
\end{table}

The results of this calculation are shown in part A of Table~\ref{t1}.
We find an almost pure $c\bar c(1 ^3P_1)$ state with mass $3467\,MeV$
which we identify with the $\chi_{c_1}(1P)$ and two states with
significant molecular admixture. One of them with mass $3865\,MeV$ is
almost a $DD^*$ molecule bound by the coupling to the $c\bar c$ states.
The second one, with mass $3936\,MeV$, is a $c\bar c(2 ^3P_1)$ with
sizable $DD^*$ component. We assign the first state to the $X(3872)$,
being the second one a candidate to the $X(3940)$.
We have also analyzed the effect of higher bare $c\bar c$ states 
finding a negligible effect on the mass and probabilities that will not
change the above numbers. 

Coexistence of the $\omega J/\psi (I=0)$ and $\rho J/\psi (I=1)$ decay 
modes strongly suggest a large isospin mixing. However the relative branching 
fraction of both modes can be misleading with respect to the absolute magnitude of the isospin mixing in $X(3872)$
 due to the phase space suppression of the $\omega J/\psi$ channel against the $\rho J/\psi$ one. In fact if we assume 
 that $X(3872)$ is a $D^0 {D^*}^0$ molecule, the ratio $\frac{B(X(3872)\rightarrow \pi^+\pi^-\pi^0J/\psi)}{B(X(3872)\rightarrow \pi^+\pi^-J/\psi)}$ would be a factor 20 smaller than the experiment due to the different phase space.

It is clear that we need charged components in the wave function but with a different weight with respect to the neutral component. This rules out the intuitive idea of the dominance of the loosely bound neutral component. The clarification of this puzzle has been nicely done in Ref~\cite{O1}.

To introduce the isospin breaking in our calculation we turn to the charge basis
instead of the isospin symmetric basis with
the transformation
\begin{eqnarray}
	|D^\pm {D^*}^\mp \rangle &=& 
	\frac{1}{\sqrt 2} \left( |D {D^*} I=0 \rangle - |D {D^*} I=1 \rangle \right)
\\
	|D^0 {D^*}^0 \rangle &=& 
	\frac{1}{\sqrt 2} \left( |D {D^*} I=0 \rangle + |D {D^*} I=1 \rangle \right)
\end{eqnarray}
writing our isospin symmetric interaction on the charged basis. We now
explicitly break isospin symmetry taking the experimental threshold difference
into account in our equations and solving for the charged and neutral components.
Of course, if we don't break it explicitly we recover our previous result as
a bound state in the $I=0$ sector. Now we get again three states being the main
difference in the $DD^*$ molecular component. The masses and channel probabilities
are shown in part B of Table \ref{t1}.
We now get a higher probability for the $D^0 {D^*}^0$ component although
the isospin 0 component still dominates with a $66\,\%$ probability and a 
$3\,\%$ for isospin 1.

\begin{figure}[t]
\begin{center}
\scalebox{1.1}{\includegraphics{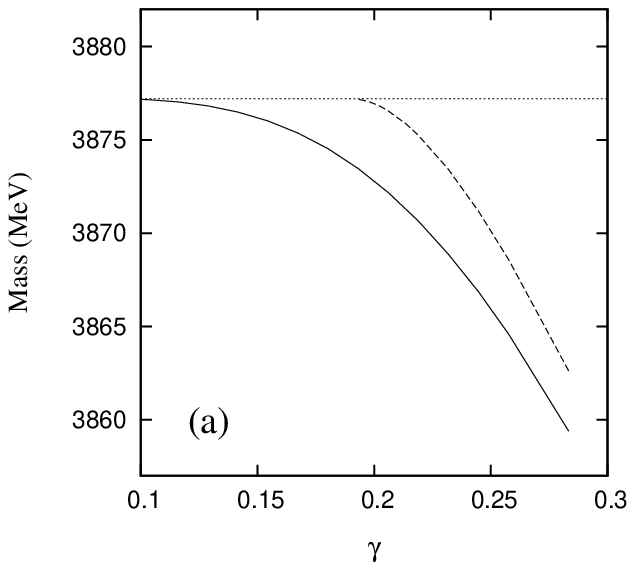}}
\scalebox{1.1}{\includegraphics{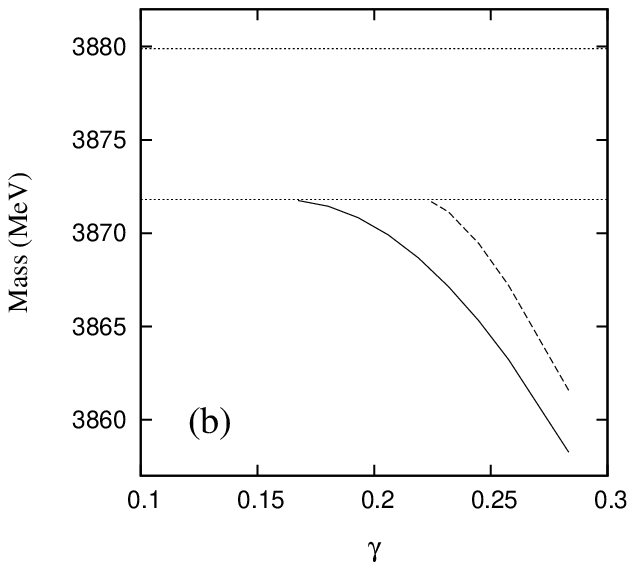}}
\caption{\label{fig1} Mass of the $X(3872)$
as a function of the strength $\gamma$ of the $^3 P_0$
model. 
The isospin symmetric
calculation is shown in figure (a) and the isospin breaking in figure (b).
Dotted lines show the threshold positions for the $DD^*$ average
in figure (a) and 
$D^0{D^*}^0$ and $D^\pm{D^*}^\mp$ in (b).
The solid lines shows the full result and
the dashed lines turning off the $DD^*$ interaction.}
\end{center}
\end{figure}

\begin{figure}[t]
\begin{center}
\scalebox{1.3}{\includegraphics{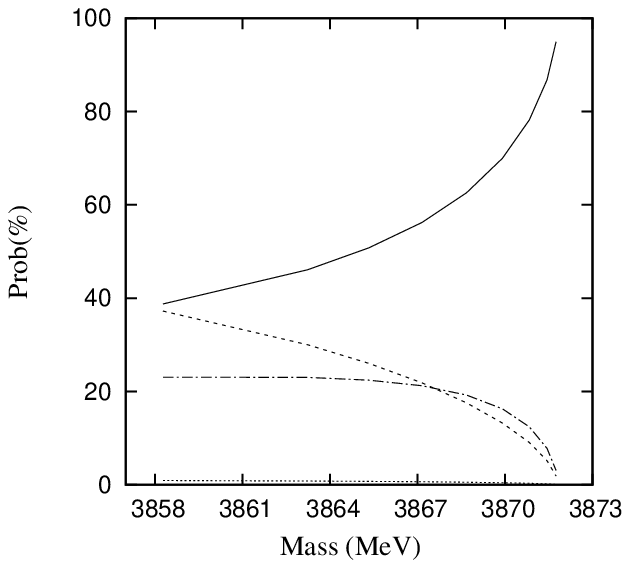}}
\caption{\label{fig2} Probability (in $\%$) of different components as a function
of the binding energy when we vary the $\gamma$ parameter of the $^3 P_0$
model. The solid line
gives the $D^0{D^*}^0$ probability, the dashed-dotted the $D^\pm{D^*}^\mp$,
the dashed the $c\bar c(2 ^3P_1)$ and the dotted the $c\bar c(1 ^3P_1)$.}
\end{center}
\end{figure}

Having in mind that the $^3 P_0$ model is probably too naive and we might be
overestimating the value of $\gamma$, we show
in Fig.~\ref{fig1} the variation of the  $X(3872)$ mass with it.
We can see that it is possible to get the experimental binding energy with a 
fine tune of this parameter. Using $0.6\,MeV$ as the binding energy we
get a value of $\gamma=0.19$, $25\,\%$ smaller than the original. 
The results are shown in part C of Table \ref{t1}.
Now the $D^0 {D^*}^0$ clearly dominates with a $83\,\%$ probability giving
a $70\,\%$ for the isospin 0 component and $23\,\%$ for isospin 1.
Of course, as the isospin breaking
is a threshold effect~\cite{r13}, it grows as we get closer 
to it as can be seen
in Fig.~\ref{fig2} where we show the probabilities of the different components 
for the state $X(3872)$.

\begin{figure}[t]
\begin{center}
\scalebox{1.1}{\includegraphics{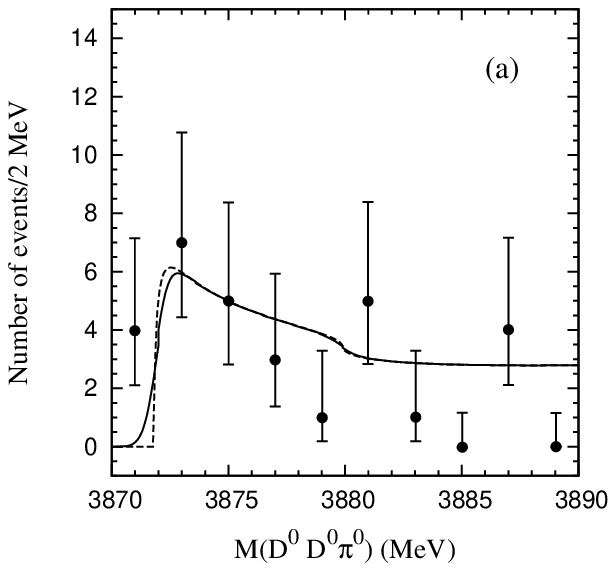}}
\scalebox{1.1}{\includegraphics{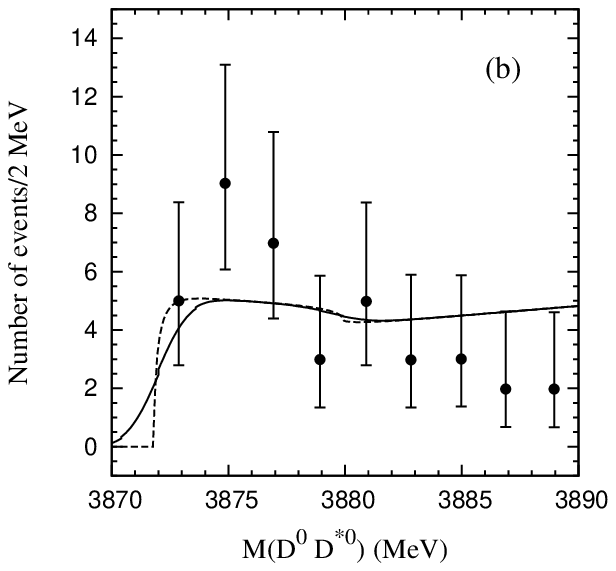}}
\caption{\label{fig3} Number of events for the decay
$B\to KD^0D^0\pi^0$ measured by Belle (a) and for
the decay $B\to KD^0D^{*0}$ measured by BaBar (b).
The solid and dashed lines shows the results from
our model with and without the resolution functions
as explained in the text.}
\end{center}
\end{figure}

\begin{figure}[t]
\begin{center}
\scalebox{1.1}{\includegraphics{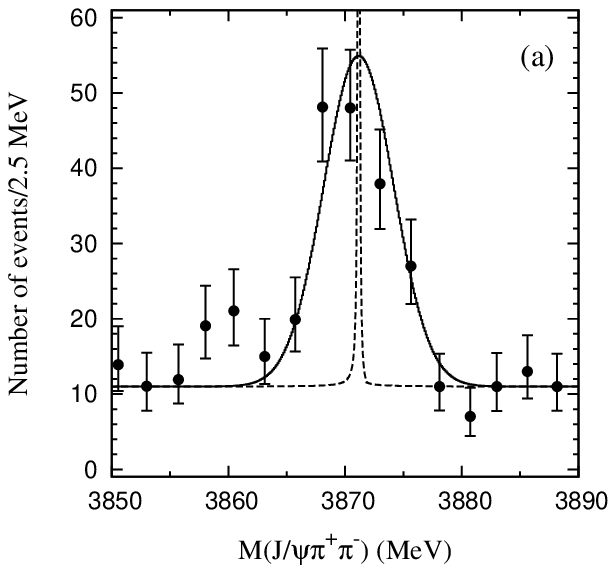}}
\scalebox{1.1}{\includegraphics{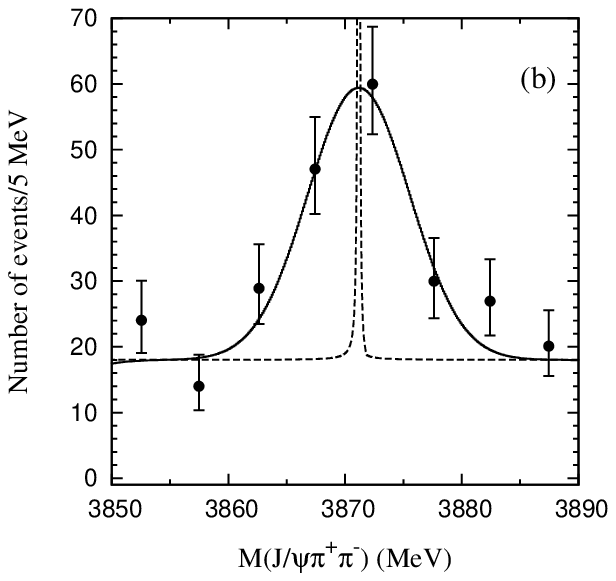}}
\caption{\label{fig4} Number of events for the decay
$B\to K\pi^+\pi^-J/\Psi$ measured by Belle (a) and 
by BaBar (b).
The solid and dashed lines shows the results from
our model with and without the resolution functions
as explained in the text.}
\end{center}
\end{figure}

In Fig.~\ref{fig3} we compare our results with the
$B\to K  D^0\bar D^{0}\pi^0$ data from Belle (a) and
$B\to K  D^0\bar D^{*0}$ data from BaBar (b).
The same comparison is done in
Fig.~\ref{fig4} for the
$B\to K  \pi^+\pi^-J/\Psi$ data from Belle (a)
and BaBar (b). 
In all figures the dashed lines shows
the results without resolution functions. The solid line
gives the result using the resolution functions as in
Ref.~\cite{Kala1}. All the resolution functions
are those given by Belle~\cite{BelleDDD} and BaBar~\cite{BaBarDJPsi} collaboration
with the exception of the BaBar $DD^*$ resolution where
we use the prescription from Ref.~\cite{Kala1}.

We find a good description of the Belle $B\to K D^0 D^0\pi^0$
data
whereas the agreement is poor in the case of
the BaBar data. It is important to notice
that in the Belle analysis the mass of the
$X$ appears as $3872\,MeV$ while in the
BaBar data the resonance is located $3\,MeV$
above. The BaBar mass value does not
coincides with the
mass of the $X$ obtained in our
calculation which may be the reason
for the disagreement.

The $B\to K  \pi^+\pi^-J/\Psi$ data are
equally well described for the Belle and 
BaBar experiments. In this case both
Collaborations give similar values for the mass of
the resonance, namely $3871.4\,MeV$, which are in
much better agreement with our result.

\section{Summary.}
As a summary, we have shown that the $X(3872)$ emerges in a constituent
quark model calculation as a dynamically generated mixed state of a $DD^*$ molecule and 
$\chi_{c_1}(2P)$. Although the $c\bar c$ mixture is less than the $10\%$ it is important to bind the molecular state.
This result is in agreement with the analysis of Ref ~\cite{g1}.
The proposed structure allows to understand simultaneously the isospin violation
showed by the experimental data and the radiative decay rates.
Furthermore, we have demonstrated that this solution explain the
new Belle data in the $D^0D^0\pi^0$ and $\pi^+\pi^-J/\Psi$ decay
modes and the $\pi^+\pi^-J/\Psi$ BaBar data. 
The original $\chi_{c_1}(2P)$ state acquires a significant
$DD^*$ component and can be identified with the $X(3940)$.

\acknowledgments
This work has been partially funded by Ministerio de Ciencia y Tecnolog\'\i a
under Contract
No. FPA2007-65748, by Junta de Castilla y Le\'on under Contract No.
SA-106A07 and GR12, 
by the European Community-Research Infrastructure Integrating
Activity ``Study of Strongly Interacting Matter'' (HadronPhysics2 
Grant no. 227431) and
by the Spanish Ingenio-Consolider 2010 Program
CPAN (CSD2007-00042).

\end{document}